\documentclass[aps,reprint,superscriptaddress,prl,floatfix]{revtex4-1}
\usepackage[utf8]{inputenc}
\usepackage[T1]{fontenc}
\usepackage{amsmath,amssymb,graphicx,color,multirow}
\graphicspath{{figures/}{supplementary/figures/}}
\usepackage[bookmarks=false]{hyperref}
\usepackage[usenames, dvipsnames]{xcolor}
\usepackage{subcaption}
\usepackage[super]{nth}
\usepackage[normalem]{ulem}
\def\STO{SrTiO${}_3$~}
\def\LSAT{(LaAlO${}_3$)${}_{0.3}$(Sr${}_2$TaAlO${}_6$)${}_{0.7}$~}

\def\RSNOx{R$_{1-x}$Sr$_x$NiO${}_2$~}
\def\LSNO{La${}_{0.8}$Sr${}_{0.2}$NiO${}_2$~}
\def\LSNOu{La${}_{0.85}$Sr${}_{0.15}$NiO${}_2$~}

\def\NSNO{Nd${}_{0.825}$Sr${}_{0.175}$NiO${}_2$~}
\def\NSNOx{Nd${}_{1-x}$Sr${}_{x}$NiO${}_2$~}

\def\RNO{$R$NiO${}_2$~}
\def\LNO{LaNiO${}_2$~}

\def\NNO{NdNiO${}_2$~}

\def\PNO{PrNiO${}_2$~}

\def\LSCO{La${}_{2-x}$Sr${}_x$CuO${}_4$~}

\def\spinglass{SG~}

\def\Tglass{\ensuremath{T_g}~}

\newcommand{\vc}[1]{\ensuremath{\mathbf{#1}}~}
\newcommand{\figref}[1]{Fig.~\ref{#1}~}

\begin{document}

\title{Spin-glass state in nickelate superconductors}

\author{David R. Saykin}
\affiliation{Stanford Institute for Materials and Energy Sciences, SLAC National Accelerator Laboratory, 2575 Sand Hill Road, Menlo Park, CA 94025, USA.}
\affiliation{Geballe Laboratory for Advanced Materials, Stanford University, Stanford, CA 94305, USA.}
\affiliation{Department of Physics, Stanford University, Stanford, CA 94305, USA.}

\author{Martin Gonzalez}
\affiliation{Stanford Institute for Materials and Energy Sciences, SLAC National Accelerator Laboratory, 2575 Sand Hill Road, Menlo Park, CA 94025, USA.}
\affiliation{Department of Materials Science and Engineering, Stanford University, Stanford, California 94305, USA.}

\author{Jennifer Fowlie}
\affiliation{Stanford Institute for Materials and Energy Sciences, SLAC National Accelerator Laboratory, 2575 Sand Hill Road, Menlo Park, CA 94025, USA.}
\affiliation{Department of Applied Physics, Stanford University, Stanford, CA 94305, USA.}

\author{Steven A. Kivelson}
\affiliation{Department of Physics, Stanford University, Stanford, CA 94305, USA.}

\author{Harold Hwang}
\affiliation{Stanford Institute for Materials and Energy Sciences, SLAC National Accelerator Laboratory, 2575 Sand Hill Road, Menlo Park, CA 94025, USA.}
\affiliation{Geballe Laboratory for Advanced Materials, Stanford University, Stanford, CA 94305, USA.}
\affiliation{Department of Applied Physics, Stanford University, Stanford, CA 94305, USA.}

\author{Aharon Kapitulnik}
\email{aharonk@stanford.edu}
\affiliation{Stanford Institute for Materials and Energy Sciences, SLAC National Accelerator Laboratory, 2575 Sand Hill Road, Menlo Park, CA 94025, USA.}
\affiliation{Geballe Laboratory for Advanced Materials, Stanford University, Stanford, CA 94305, USA.}
\affiliation{Department of Physics, Stanford University, Stanford, CA 94305, USA.}
\affiliation{Department of Applied Physics, Stanford University, Stanford, CA 94305, USA.}

\date{\today}

\begin{abstract}
    Magneto-optical measurements in \LSNO and \NSNO reveal an intriguing new facet of infinite-layer nickelate superconductors: the onset of spin-glass behavior at a temperature far exceeding the superconducting critical temperature $T_c$. This discovery sharply contrasts with copper oxide superconductors, where magnetism and superconductivity remain largely exclusive. 
    Moreover, the magnitude and onset temperature of the polar Kerr effect in \NSNO 
    fabricated on \STO and \LSAT substrates differ dramatically, while $T_c$ does not.
\end{abstract}
\pacs{NaN}

\maketitle

\section{Introduction}
It has been a long quest to find a system analogous to the cuprates, in which a parent compound can be doped to reveal high-$T_c$ superconductivity \cite{Keimer2015}.
A typical high-$T_c$ cuprate is characterized by a stack of $n$-CuO$_2$ planes each separated by an alkaline-earth element A$^{2+}$   (A=Ca,Sr,Ba) with additional intervening ``charge-reservoir'' layers that separate the stacks (see e.g. \cite{Park1995}).
This structure implies an infinite-layer cuprate structure (where $n\to\infty$) ACuO$_2$ \cite{Siegrist1988}, which is a strongly correlated antiferromagnetic charge-transfer insulator \cite{Vaknin1989,Scalettar1991}.

However, its counterpart nickelate system \RNO ($R$ = La, Nd, Pr) is either weakly insulating \cite{Ikeda2013} or superconducting \cite{Parzyck2024,Sahib2024} and lacks long-range antiferromagnetic order \cite{Hayward2003}, while exhibiting a spin-glass behavior \cite{Lin2022}. 
This is despite sharing the same $P4/mmm$ space group and the 3d$^9$ configuration in its electronic structure. With the discovery of superconductivity in Sr-doped \NNO with $T_c$ in the range of $9$ to $15$ K \cite{Li2019}, it became clear that the set of a priori expectations for a parent material to exhibit superconductivity needs to be revisited \cite{Botana2020}. 

\begin{figure}
    \centering
    \includegraphics[width=\linewidth]{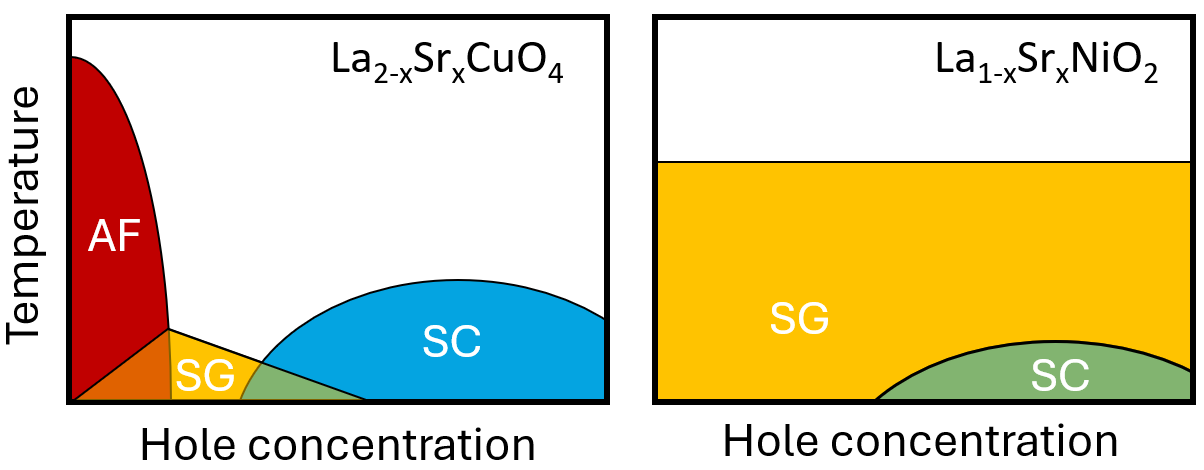}
    \caption{Schematic phase diagrams
    showing spin-glass order in cuprates and nickelates. 
    \textbf{Left:} The copper oxide superconductor \LSCO 
    exhibits AFM order for $0 \leq x \lesssim 2 \%$, and SC 
    for $5 \% \lesssim x \lesssim 25 \%$. The SG state starts within the AFM phase and apparently leaks into the SC phase \cite{Niedermayer1998,Stilp2013}. 
    \textbf{Right:} Nickelate superconductors have a SG phase
    that is present in a wide range of Sr doping,  such that the SC dome ($10 \% \lesssim x \lesssim 25 \%$) is located deep inside 
    the SG phase. The exact dependence of SG order on doping concentration is yet to be determined.}
    \label{fig:phase_diagram}
\end{figure}

In many cuprates, doping typically induces a spin-glass (SG) phase, which bridges the loss of antiferromagnetic (AFM) long range order and the emergence of superconductivity (SC) \cite{Aharony1988}. For example, a thorough study \cite{Julien2003} of the magnetic phase diagram of La$_{2-x}$Sr$_x$CuO$_4$, a single-layer doped cuprate, revealed that \spinglass order persists far into the superconducting region with a glass transition temperature \Tglass that is anti-correlated with the superconducting $T_c$. 
This is also true for thin films, where the glass transition was shown to be larger, approximately twice as large as the bulk superconducting transition \cite{Stilp2013}. 
While \Tglass seems to diminish towards optimal doping, it seems to show a peak at dopant concentration $x \approx 1/8$, which was previously shown to exhibit a weakened superconducting state. Excluding that last feature, \figref{fig:phase_diagram} is the accepted phase diagram for the spin glass phase in cuprates. 
At low doping ($x \lesssim 2\%$) the transition can be ascribed to freezing of the spins of the doped holes into a \spinglass state which is superimposed on the preexisting 3D AFM long-range order of the Cu$^{2+}$ spins \cite{Niedermayer1998, Stilp2013, Tallon1997, Julien2003}. 
With increased doping ($x \gtrsim 2\%$), stronger frustration of the AFM environment yields a lower-temperature \spinglass phase associated with freezing of spin fluctuations \cite{Chou1995,Hasselmann2004, Andersen2011}. 

While the magnetic structure of infinite-layer nickelate superconductors is yet to be fully understood, particularly as it affects superconductivity, its most salient features are the absence of long--range AFM ordering and freezing of spins into a glass phase even in the parent compound \cite{Lin2022,Ortiz2022,Puphal2023}. 
In the cuprates where magnetism originates from copper spins in the CuO$_2$ planes, which is also the starting point for models of \spinglass state.
By contrast, glassy dynamics observed in polycrystalline \LNO samples were attributed to the presence of subtle local oxygen disorder in the form of remaining apical oxygen \cite{Lin2022}.

\begin{figure}
    \centering
    \includegraphics[width=\linewidth]{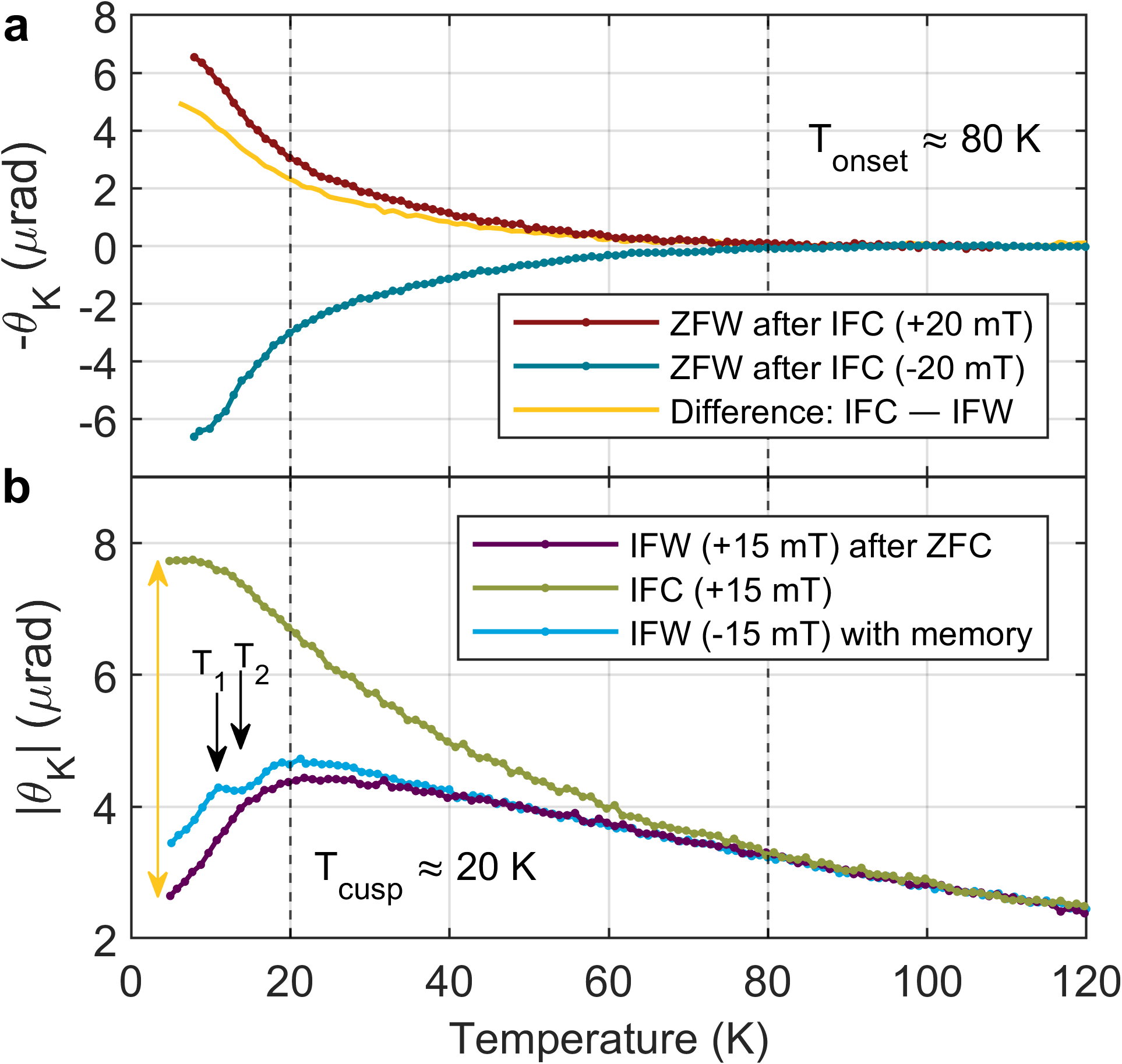}
    \caption{Kerr signal in \LSNO. %
        Vertical dashed lines indicate the onset temperature $T_\text{onset}$ and a cusp/peak feature temperature $T_\text{cusp}$.
        \textbf{a} 
        The Kerr signal recorded in a zero magnetic field warmup (ZFW) after cooling in $B=\pm20$ mT. %
        The difference between IFC in +15 mT and IFW at the same field following a ZFC is shown in yellow.
        \textbf{b} Kerr signal value 
        measured during a warmup in $+15$ mT after a zero-field cooldown (purple), and during cooldown in same field (green).
        The cyan curve demonstrates memory effect (see main text).
        }
    \label{fig:MOKE_LSNO_on_STO}
\end{figure}

\begin{figure}
    \centering
    \includegraphics[width=\linewidth]{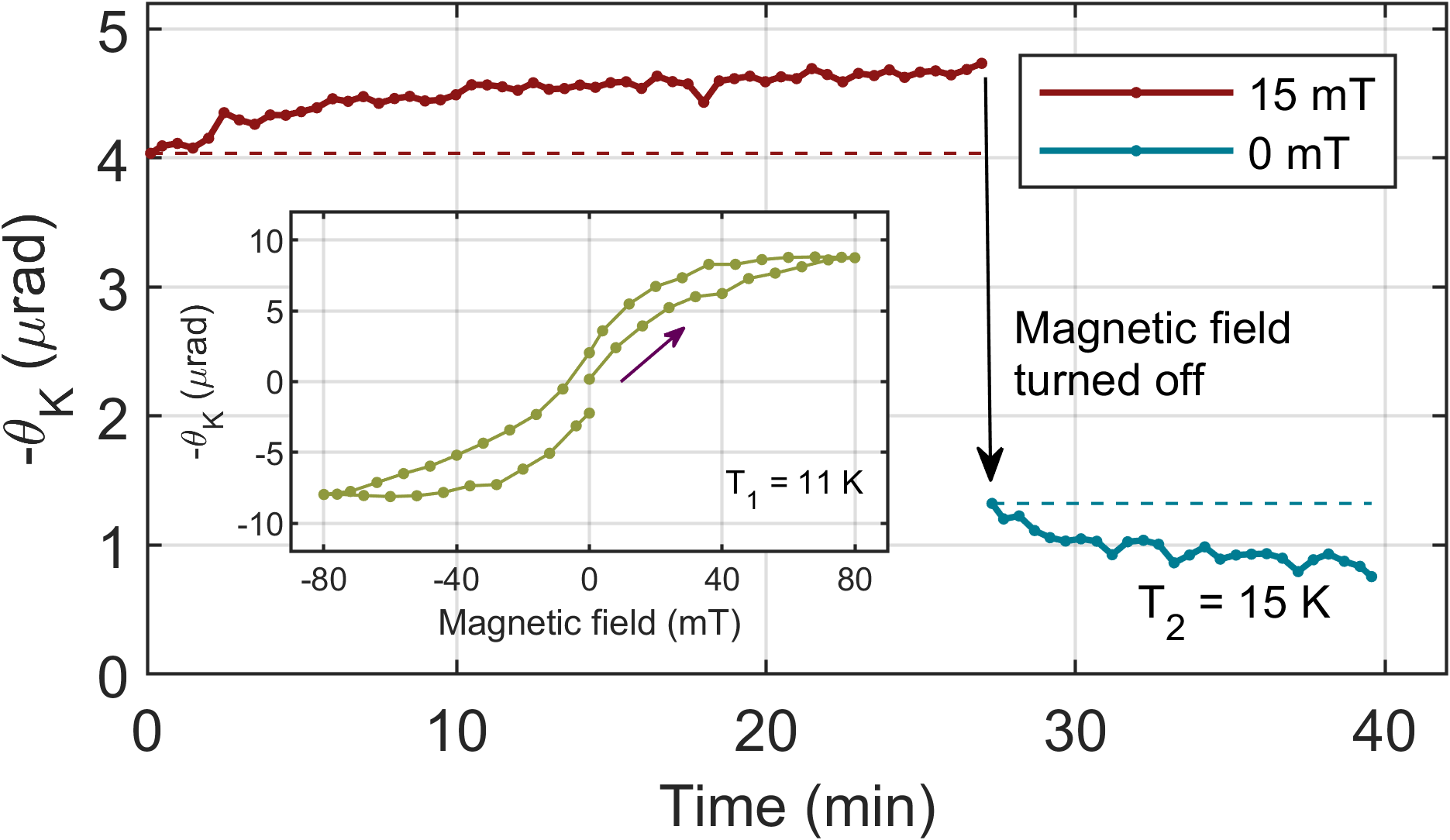}
    \caption{Slow time saturation and relaxation of magnetization in \LSNO at $T_2 = 15$ K. The magnetic field is abruptly turned on 
    at $t=0$ and then turned off after $t = 27$ minutes. %
    Horizontal dashed lines are guide for the eye. %
    \textbf{Inset:} Hysteresis loop recorded at $T_1 = 11$ K. Arrow indicates the start of the loop. Field was ramped in steps of $8$ mT each $30$ seconds.}
    \label{fig:MOKE_LSNO_on_STO_time}
\end{figure}

\begin{figure*}
	\centering
	\includegraphics[width=.45\textwidth]{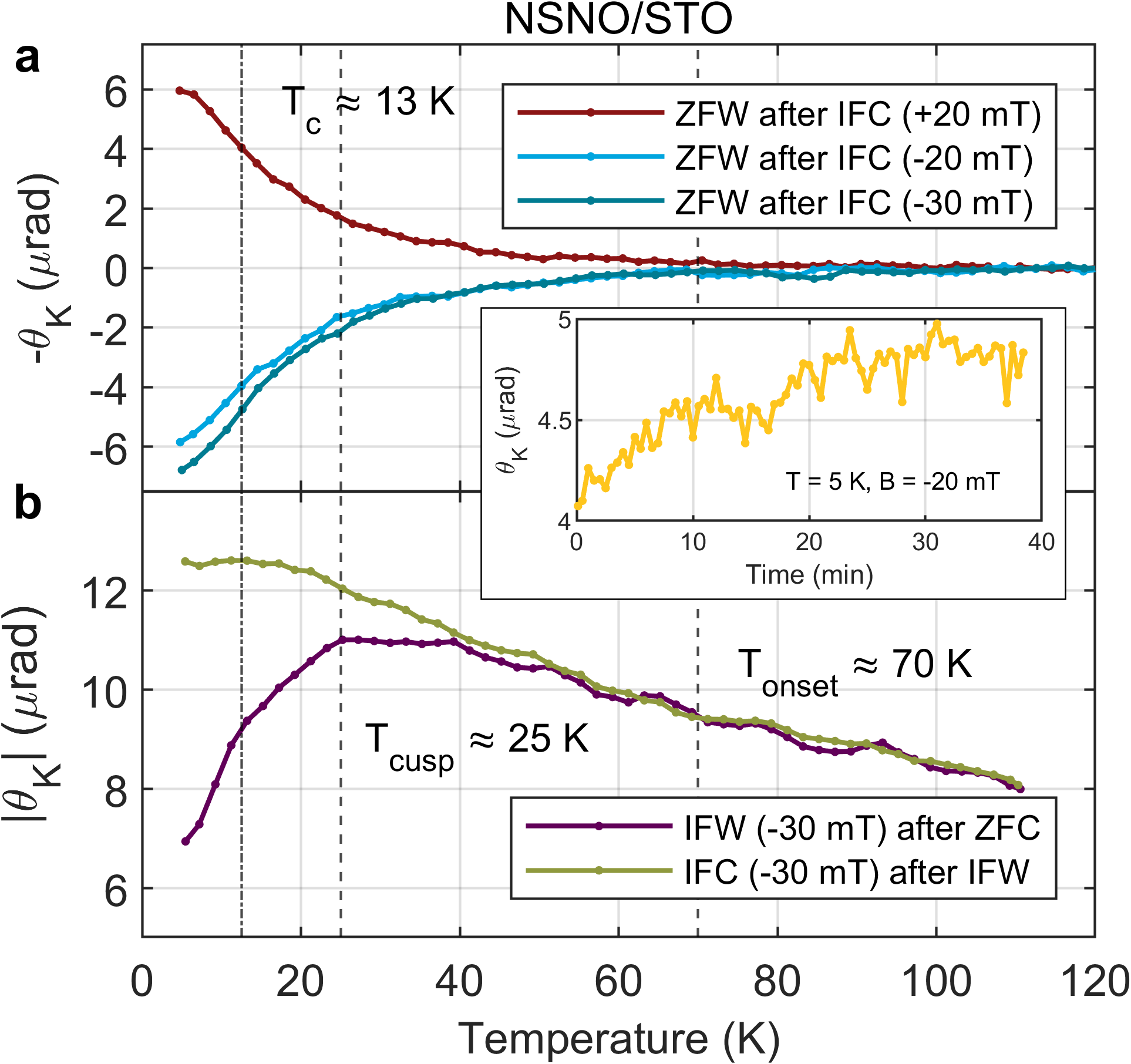}
    \qquad\quad
    \includegraphics[width=.45\textwidth]{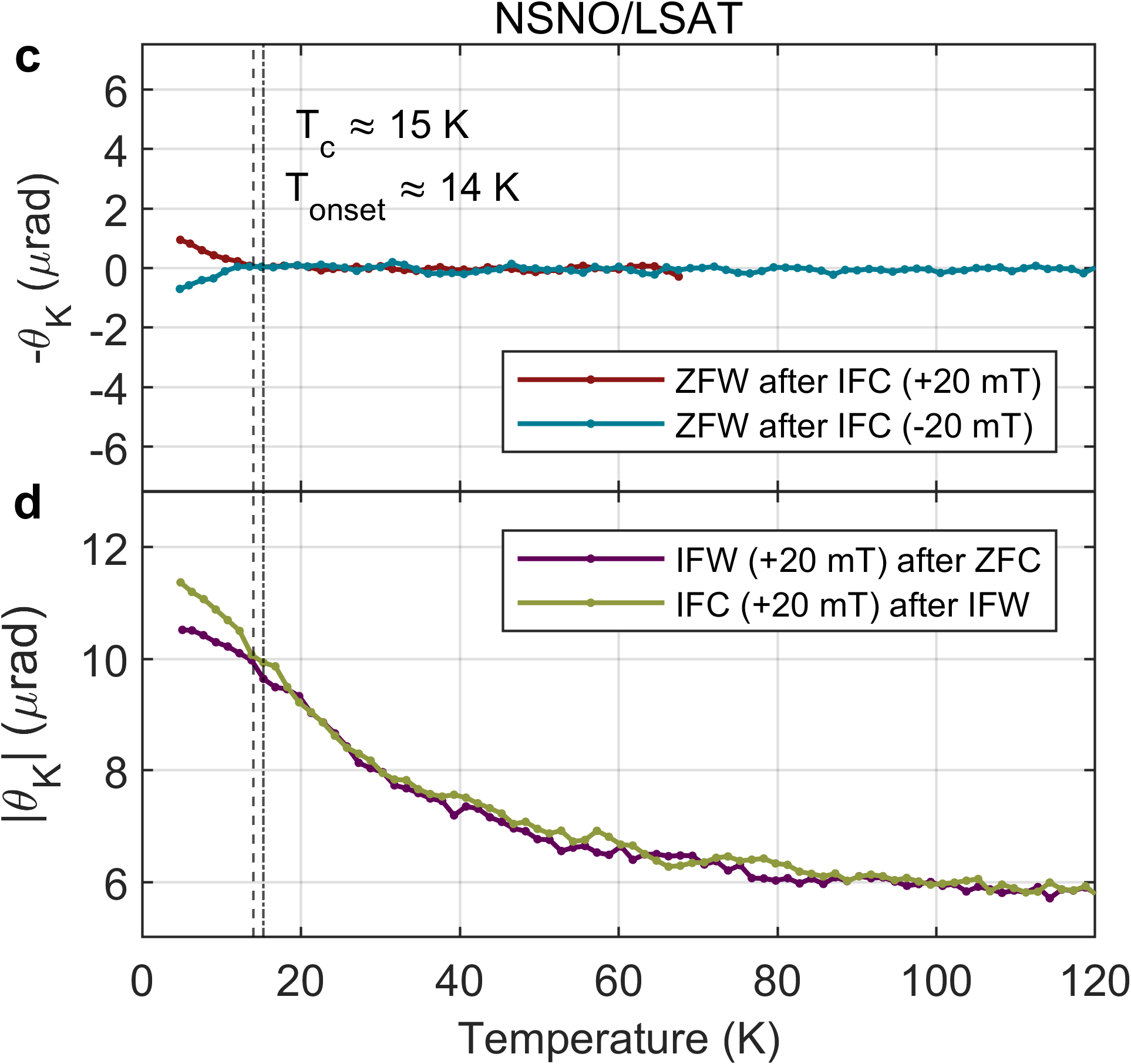}
	\caption{Kerr signal in \NSNO.
        \textbf{a,b} NSNO on STO substrate.
        \textbf{a} Data, taken during a zero-field warm-up (ZFW) preceded by cooldown from $150$ K in magnetic field.
        \textbf{b} Kerr rotation in nickelate in the presence of $-30$ mT magnetic field taken during the warmup after zero-field cooldown and during the in-field cooldown.  
       \textbf{c,d} NSNO on LSAT substrate. 
        Measurement protocol in \textbf{c} and \textbf{d} is the same as in \textbf{a} and \textbf{b} respectively.}
	\label{fig:MOKE_NSNO}
\end{figure*}

Because superconductivity has so far only been detected 
in thin films nickelates, their associated magnetic states 
have not yet been established.
A comprehensive muon spin-relaxation ($\mu$SR) study of \RSNOx by Fowlie \emph{et al.} \cite{Fowlie2022} discovered intrinsic magnetism, which gradually onsets in the $100$-$150$ K temperature range across the rare earth ($R$) series, and co-exists with superconductivity.
These authors speculated that the observed glassy magnetic phase is intrinsic in nature and
reflects the AFM coupling between the Ni$^+$ spins.
On the other hand, scanning superconducting quantum interference device (SQUID) microscopy has picked up an inhomogeneous ferromagnetic background in \LSNOu slightly above the superconducting temperature, which has been attributed to extrinsic NiO$_x$ particles on the interface between the nickelate and the \STO capping layer \cite{Shi2024}.

Moreover, measurements of the magnetic excitations in \NSNOx using resonant inelastic x-ray scattering (RIXS) indicate strong spin-1/2 AFM magnons \cite{Lu2021}, 
while nuclear magnetic resonance (NMR) measurements observe short-range glassy antiferromagnetic fluctuations \cite{Cui2021}.  
These dissimilar observations call for a better understanding of the relationship between magnetism and superconductivity in infinite-layer nickelate superconductors.

In this paper we present a comprehensive study of the magnetic state of doped infinite layer nickelates through highly sensitive measurements of the  magneto-optical polar~Kerr (MOKE) and Faraday (MOFE) effects using zero area-loop Sagnac interferometers (ZALSI) \cite{Xia2006}. 
Our work provides direct evidence 
of a spin-glass magnetic state 
in infinite-layer nickelate superconductors through observations of: 
{\it i}) Irreversible magnetic susceptibility, 
{\it ii}) Slow dynamics and aging,  
{\it iii}) memory effect. 
Focusing on doping levels close to optimal doping, 
we further demonstrate the robustness of superconductivity  in the presence of a glass phase with $\Tglass > T_c$.

\section{Results}

The detailed design of ZALSI is described in the Methods section and Supplementary Information \cite{supp}.
Most importantly, the Kerr angle is a direct probe of magnetization since it is proportional to Hall conductivity at the frequency of light \cite{Oppeneer1999,Kapitulnik2015,Fried2014}. 
\begin{align}
    \theta_K 
    \propto \sigma_{xy}(\omega) 
    \propto \pm M_z
\end{align}
Since time-reversal symmetry breaking is a necessary condition for a finite Kerr effect \cite{Halperin1992}, we will refer to polar Kerr rotation simply as the \emph{magnetic signal}.


Figures \ref{fig:MOKE_LSNO_on_STO} and \ref{fig:MOKE_LSNO_on_STO_time} 
show evidence for spin glass behavior in \LSNO (LSNO).
Figure \ref{fig:MOKE_LSNO_on_STO} demonstrates the presence of spontaneous magnetization, which gradually emerges in LSNO below $T_\text{onset} \simeq 80$ K.
When the sample is field-cooled in a small field of $\pm20$ mT and then the field is turned off at the base temperature $T_\text{base} \sim 5$ K, any remanent Kerr rotation is evidence for aligned magnetic moments in the material. Reversing the direction of the training field leads to the reversal of the sign of the Kerr signal, which confirms the magnetic nature of the optical signal. The observed smooth onset of the magnetic signal is characteristic for glassy systems.

In-field cooldown and warmup measurements (bottom panel of Fig.~\ref{fig:MOKE_LSNO_on_STO}) probe magnetic susceptibility, which appears to be irreversible: 
following a zero-field cooldown (ZFC), magnetic moments are frozen and cannot be rotated by a weak external field, while during the in-field cooldown  (IFC), moments are aligned with the field. The difference between these two measured susceptibilities (IFW and IFC) is roughly the same as the remanent magnetization extracted during the measurement with ZFW protocol as shown by the yellow line on the top panel of the \figref{fig:MOKE_LSNO_on_STO}.

Irreversible behavior of magnetic susceptibility and gradual onset of magnetization are known properties of spin-glass; however these could also be explained with domain physics alone. Thus, to verify the \spinglass nature of magnetic state in nickelate superconductors, we further probe dynamical manifestations of \spinglass phase.

After cooling the sample in zero-field from room temperature to $T_\text{base}$, we heat it up to $T_2 = 15$ K, turn on a weak external field $B = +15$ mT and record Kerr signal as a function of time. As evident from Figure \ref{fig:MOKE_LSNO_on_STO_time}, magnetization slowly increases on the scale of minutes; this property of spin-glass is known as \emph{aging}. 
Once the field is turned off, the magnetization slowly relaxes to some finite but smaller value on a similar time scale. The observed time-dependence of magnetic susceptibility is characteristic of glass-like spin dynamics and is comparable to analogous experiments performed on bulk polycrystalline \LNO samples \cite{Lin2022, Ortiz2022}.

From $T_2$ we cool the sample down to $T_1 = 11$ K and perform a hysteresis loop by sweeping the field to $+80$ mT, then to $-80$ mT and back to zero. 
The resulting hysteresis curve is depicted in the inset of \figref{fig:MOKE_LSNO_on_STO_time}. 
Finally, we cool down to a base temperature of $T_0 = 5$ K, 
turn on an opposite field $B = - 15$ mT and record the Kerr signal during in-field warmup, 
which exhibits a relatively sharp, 
non-monotonic features at $T_1 = 10$ and $T_2 = 15$ K 
(see bottom panel of Figure \ref{fig:MOKE_LSNO_on_STO}). 
This unusual effect can be explain as  a manifestation of the ``memory effect,'' 
which happens because the systems retains information about spin dynamics that was happening at $T_1$ and $T_2$ prior to cooldown. Above $T_2$ the magnetic susceptibility recovers the same trend that was recorded during the previous in-field warmup after zero-field cooldown.
The hierarchic structure of the free-energy landscape with many local minima leads to apparent "memory" about the state from which the system was cooled down \cite{Dotsenko1993}.

\begin{figure}
    \centering
    \includegraphics[width=\linewidth]{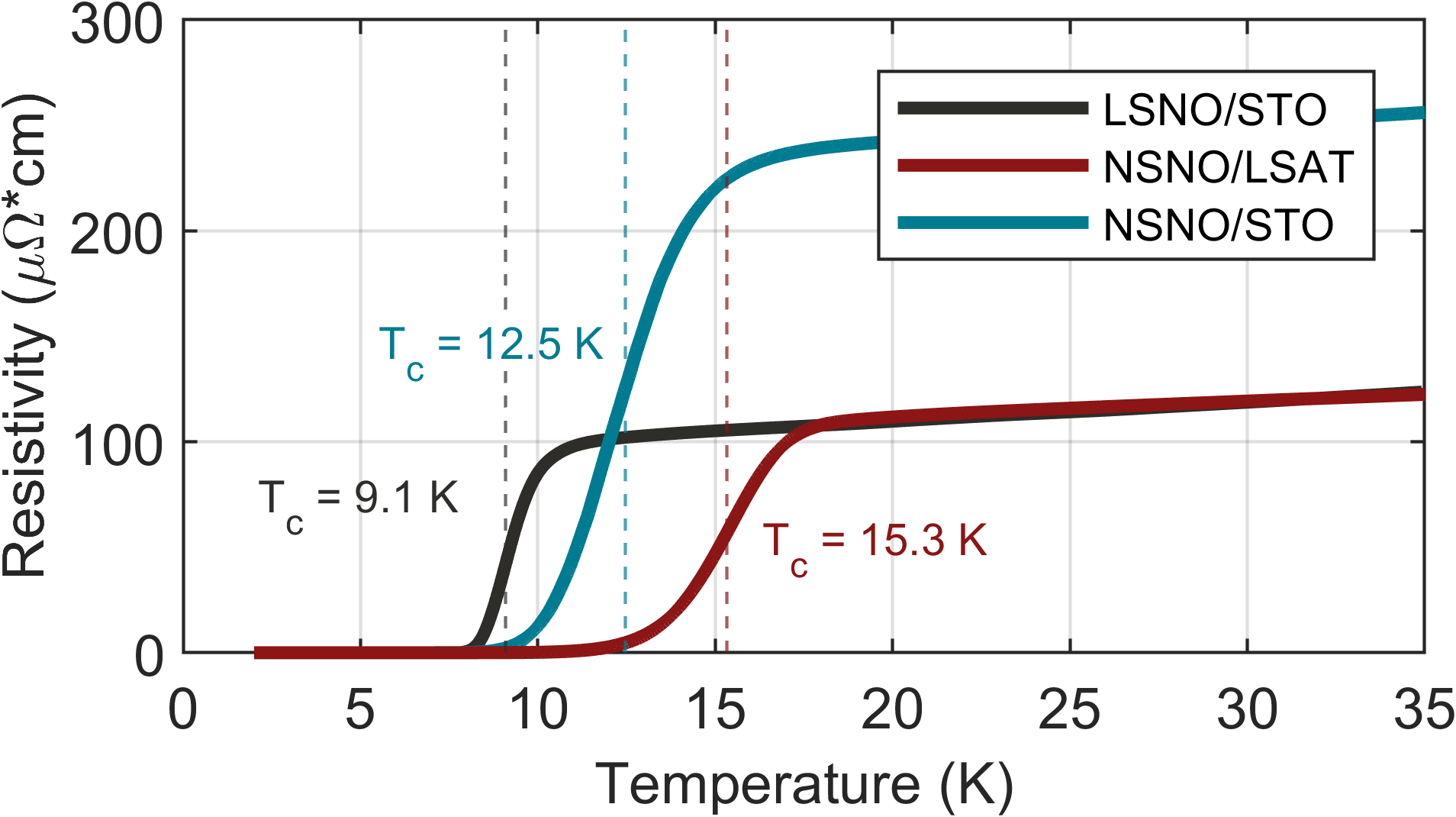}
    \caption{Resistivity of LSNO, NSNO/STO, and NSNO/LSAT. Superconducting transition temperatures $T_c$ (indicated with dashed lines) were determined as the mean of inflection point temperature and half-normal point temperature.}
    \label{fig:resistivity}
\end{figure}

Finally, we direct our attention to measurements of \NSNO (NSNO) samples prepared on \STO and \LSAT substrates, abbreviated as STO and LSAT respectively (\figref{fig:MOKE_NSNO}). It was shown that nickelate films prepared on LSAT substrate have higher crystallinity compared to the similar films fabricated on STO substrate \cite{Lee2023,Ren2023}. This is evident from the reduction in Ruddlesden-Popper-type stacking faults, improved metallicity, and an increase in the superconducting critical temperature (see Fig.~\ref{fig:resistivity}) when compared to NSNO on STO \cite{Lee2020}.

The irreversible trend in Kerr susceptibility, together with the aging effect, undoubtedly demonstrates glassy nature of magnetic order in \NSNO.
Furthermore, the data explicitly demonstrates a significant difference in the strength of magnetic signal between the two NSNO samples grown on different substrates. 
Specifically, the magnitude of the zero-field Kerr signal at base temperature drops from about $6$ $\mu$rad to about $1$ $\mu$rad, while onset temperature is decreased to $T_\text{onset} \approx 14$ K in NSNO/LSAT sample compared to $T_\text{onset} \approx 70$ K in the NSNO/STO. 
In fact, magnetic signal in NSNO/LSAT is so small, that we're unable to demonstrate aging phenomenon in this sample. Still, irreversible susceptibility and gradual onset of magnetic signal in zero-field measurements are indicative of a spin-glass state. We also note that in NSNO/LSAT samples we observe only a single characteristic temperature, at which the signal onsets, but we do not see any cusp-type features in the susceptibility data.

\section{Discussion}
Our results of the dynamics of the magnetic state of doped thin films nickelets are in good agreement with the magnetic susceptibility measurements of the polycrystalline parent compound \LNO \cite{Lin2022,Ortiz2022}, which demonstrated similar hallmarks of SG phase: irreversible susceptibility, aging, and a memory effect. 
The authors of \cite{Lin2022} have also identified two characteristic temperatures associated with \spinglass behavior in the AC susceptibility of \LNO: cusp-like peak feature around $T_\text{cusp} \simeq$ 13 K and an onset of frequency-dependent susceptibility at $T_\text{onset} \simeq $ 85 K. Approximate typical temperatures are summarized in the Table~\ref{tab:transition_temperatures}. Based on this observation, they surmised the possible existence of two disorder mechanisms at play with different characteristic energy scales. 

However, it is challenging to estimate the exact SG emergence temperature due to the gradual nature of the transition, based on $T_\text{cusp}$ and the magnitude of the splitting between susceptibility curves, magnetic signal in the thin films on the STO substrate appears to be comparable to the signal in the (undoped) nickelate crystals \cite{Lin2022}, which generally have lower crystallinity \cite{Li2020}. This similarity appears to be even more bizarre, when it is taken into account that polycrystalline samples are reported to have no nickel impurities, while $\sim 37$ nickel particles per $\mu$m$^2$ have been reported in the thin-film samples \cite{Shi2024}. In any case, the onset temperature of the \spinglass phase does not depend strongly on the rare earth ion or Sr doping, which permits us to surmise the phase diagram shown in \figref{fig:phase_diagram}.

Importantly, present magneto-optical measurements agree with the $\mu$SR study \cite{Fowlie2022}, conducted on the same \LSNO sample as measured in the current work. We speculate that the spin-glass magnetic order detected with ZALSI is responsible for the change in muon spin relaxation rate between $T=100$ K and $T = 5$ K reported in Ref. \cite{Fowlie2022}. 
Since $\mu$SR confirm the bulk nature of the magnetic signal, we conclude that the glassy spin dynamics is not caused by extrinsic factors such as Ni clusters on the interface with the capping layer, but instead arise due to intrinsic mechanisms. However, in contrast to $\mu$SR we can exclude the AFM interaction as the origin of the observed glassy magnetic state, since compensated AFM ordered moments would not produce any Kerr rotation.

Finally we comment on the large reduction in strength and characteristic temperature of the magnetic signal between the NSNO samples that are fabricated on STO and LSAT substrates. 
A relatively moderate increase by 20 \% in superconducting temperature (see \figref{fig:resistivity}), indicates that the mechanism for ``static'' magnetism in nickelates is weakly related to superconductivity. Instead, it appears that the overall improvement in the sample's crystallinity is responsible for the suppressed \spinglass state. 
Furthermore, as evident from the RIXS study of \PNO, the magnon spectrum does not change between samples prepared on STO or LSAT substrates \cite{Gao2024}. 

\begin{table}
    \begin{tabular}{ |c||c|c|c|  }
        \hline
        Sample 			& 	$T_c$	&	$T_\text{cusp}$		&	$T_\text{onset}$	\\
        \hline\hline
        \LNO (polycrystal) \cite{Lin2022}	&	-	&	13	&	85	\\\hline
        \NNO (polycrystal) \cite{Lin2022}	&	-	&	6	&	85	\\\hline
        \LSNO / STO		&	9.1		&	20	&	80	\\\hline
        \NSNO / STO		&   12.5	&	25	&	70	\\\hline
        \NSNO / LSAT	&	15.3	&	-   &   14	\\\hline
    \end{tabular}
    \caption{Summary of the transition temperatures (in Kelvins) for different samples: undoped crystalline materials from Ref. \cite{Lin2022} and optimally doped superconducting thin films.}
    \label{tab:transition_temperatures}
\end{table}

\subsection{Conclusions}
The main results of the present work are magneto-optical measurements of superconducting infinite-layer nickelates \LSNO and \NSNO, which clearly demonstrate the presence of a spin-glass phase coexisting with superconductivity and persisting up to temperatures \Tglass much higher than superconducting critical temperature $T_c$.
The dramatic change in magnetic signal and onset temperature between \NSNO samples grown on \STO and \LSAT substrates indicates the importance of the sample's crystallinity  on the disorder magnitude in the exchange interaction energy.  On the other hand, the relatively smaller effect this change has on the superconducting $T_c$ suggests that there is no simple, direct connection between SG order and the mechanism of superconductivity in this material - and possibly, by analogy - in the cuprates as well.

\section{Methods}
The ZALSI apparatus has a unique symmetry-based design that enables the measurement of the Kerr angle $\theta_K$ with sub-microradian resolution. ZALSI has been used to detect time-reversal symmetry breaking (TRSB) in superconductors \cite{Xia2006Sr2RuO4, Xia2009, Schemm2014, Hayes2021, Wei2022} and thin film ferromagnets \cite{Xia2009SrRuO3}. It was recently demonstrated that ZALSI is sensitive exclusively to TRSB and does not suffer from optical activity induced by non-magnetic phases such as charge order \cite{Saykin2023,Farhang2023,Farhang2025}. 

We use continuous superluminescent diode light source operating at $\lambda = 1550$ nm SLED, which corresponds to $\omega \simeq$ 0.8 eV. Additionally, we have repeated selected measurements with $830$ nm ($1.5$ eV) ZALSI, and found qualitatively similar results. Notably, the Kerr rotation from the LSNO at $830$ nm is reversed in sign and has a smaller magnitude compared to signal detected at $1550$ nm. 
\begin{align}
    \theta_K^{\lambda = \text{830 nm}} &\propto + M_z,   \\
    \theta_K^{\lambda = \text{1550 nm}} &\propto - M_z.
\end{align}
This behavior is, in fact, similar to elemental nickel, where Kerr angle is known to reverse sign at $\omega \approx 1$ eV \cite{Buschow1988ch5}.
Moreover, for in-field measurements paramagnetic-like contributions from the substrate is much larger at $830$ nm. This makes longer wavelengths more suitable for studying magnetism in nickelate superconductors.

Overall, magneto-optical Kerr effect measurements are perfectly suited for studying magnetization in nickelate thin-film superconductors, since they allow for local bulk probe (beam spot size $\approx 25$ $\mu$m) of magnetization in a wide range of temperatures.
However, due to tiny absolute magnitude of the \spinglass signal of $\sim 5$ $\mu$rad, this measurement is beyond the reach of conventional MOKE setups, whereas ZALSI is able to resolve such signals with ease owing to its extremely high sensitivity of $100$ nrad/$\sqrt{\text{Hz}}$.
See the Supplementary Material \cite{supp} for more details on relationship between ZALSI MOKE signal and magnetization, its dependence on wavelength, and substrate contributions.

\subsection{Acknowledgments}
We thank Daniel S. Fisher for the insightful discussions.
Work at Stanford University was supported by the U.S. Department of Energy, Office of Science, Basic Energy Sciences, Division of Materials Sciences and Engineering, under Contract DE-AC02-76SF00515.

\bibliography{bib/kerr, bib/nickelates, bib/cuprates, bib/magnetism}

\newpage

\appendix*
\setcounter{equation}{0}\renewcommand{\theequation}{A.\arabic{equation}}

\tableofcontents

\section{Samples}\label{app:samples}
The nickelate thin films investigated in this study were synthesized via a two-step process involving pulsed laser deposition of the perovskite precursor material $R_{1-x}$Sr$_x$NiO$_3$ with rare-earth elements $R$ = La, Nd, followed by topochemical reduction into the infinite layer phase  $R_{1-x}$Sr$_x$NiO$_3$. The thin films were grown on either LSAT (001) or STO (001) single crystal substrates. The substrates were pre–treated with acetone/isopropyl alcohol ultra–sonication. The \STO substrates also require a high-temperature anneal at 900 \textdegree C for 30 minutes in an oxygen partial pressure of 2 $\times$ 10$^{-6}$ Torr. The growth conditions for all samples are otherwise equivalent.

The nickelate films were deposited at 570 \textdegree C in an oxygen partial pressure of 0.15 - 0.2 Torr.  The resulting perovskite precursor films were then wrapped in aluminum foil and reacted with 0.15 g CaH$_2$ powder within a vacuum-sealed Pyrex glass tube (< 5 mtorr). The tube was heated to a temperature of 240 - 320 \textdegree C and annealed until a complete structural transition into the infinite layer phase was achieved. Further details on the fabrication process are described in the Supplementary Materials of Ref. \cite{Fowlie2022}, and the Extended figures of Ref. \cite{Lee2023} for the \LSNO  and \NSNO films, respectively.

\begin{figure}[h]
    \centering
    \includegraphics[width=\linewidth]{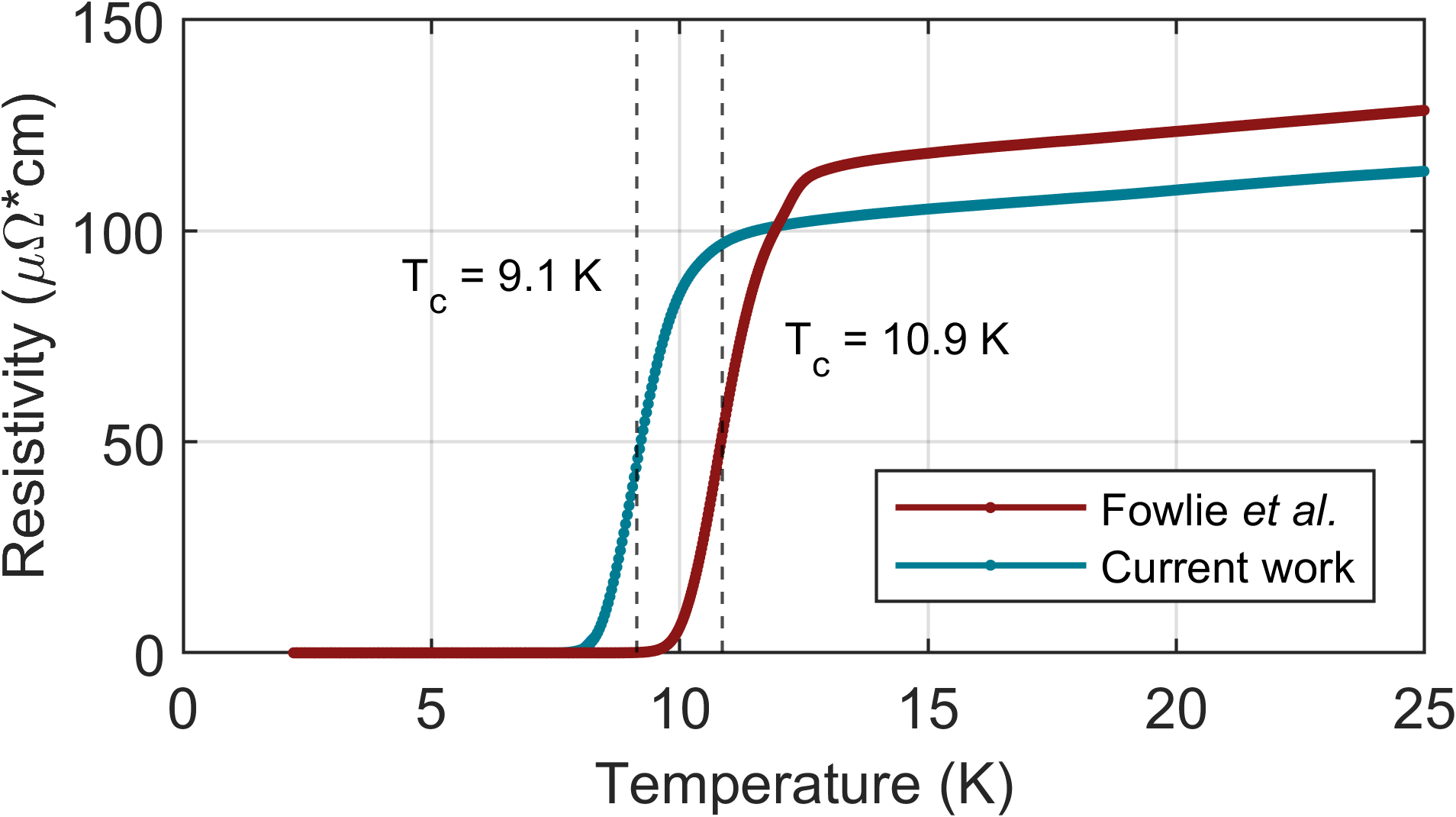}
    \caption{
        \textbf{Resistivity of \LSNO.} Resistivity of LSNO-on-STO sample measured in the present study compared to the same measurement done at the time of publication of Ref. \cite{Fowlie2022}. Superconducting transition temperatures $T_c$ were determined as the mean of inflection point temperature and half--normal point temperature.
    }
    \label{fig:LSNO_resistivity}
\end{figure}

The \LSNO sample studied in this work is one of eight $2.5 \times 5$ mm pieces constituting the mosaic of samples measured by Fowlie et al \cite{Fowlie2022}. The \LSNO sample was was grown on \STO substrate with a reduced phase thickness of 7.7 nm, and includes a 13.6 nm \STO capping layer. Resistivity measurements of the sample were repeated right after the magneto–optical measurement were conducted, and reduction of Tc due to slow degradation with time was observed (see \figref{fig:LSNO_resistivity} ) compared to the state of the sample at the time of $\mu$SR measurements \cite{Fowlie2022}. 

The \NSNO samples were grown specifically for this study, and feature a reduced phase thickness of approximately 5 nm with a 1.5 nm \STO capping layer. Resistivity of all samples is shown in Fig.~5 of the main text.

\begin{figure*}[!t]
    \centering
	\includegraphics[width=.67\textwidth]{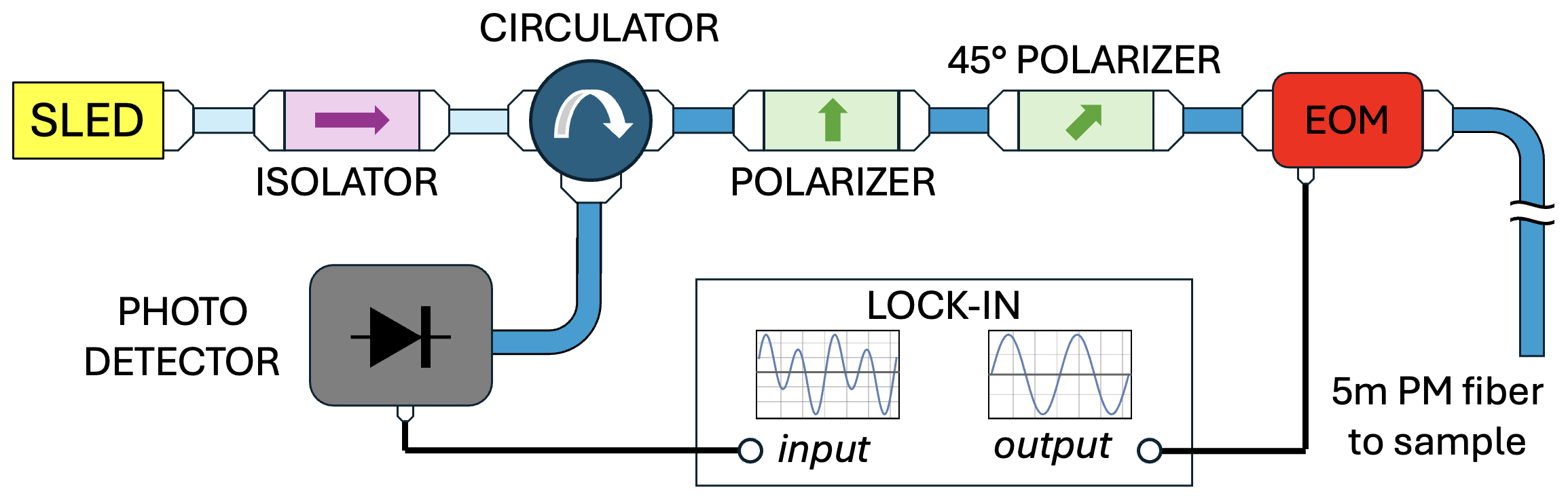}
    \qquad
    \includegraphics[width=.26\textwidth]{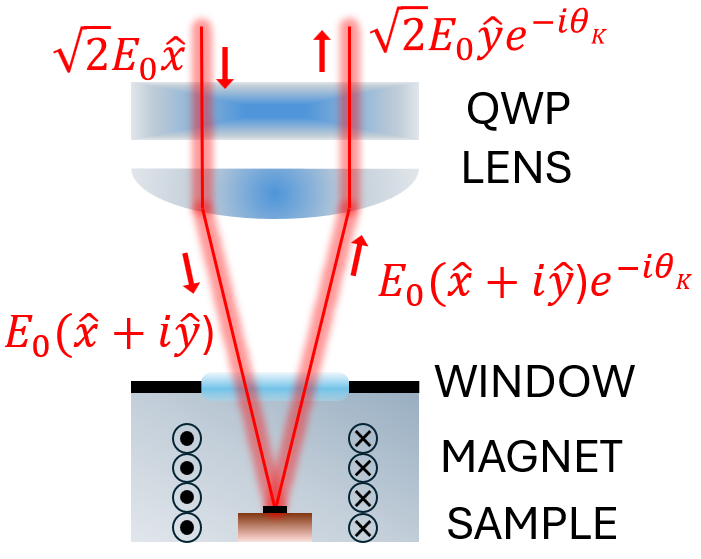}
	\caption{
        \textbf{ZALSI.} Schematics of the zero-area-loop fiber-optic Sagnac interferometer.
        \textbf{Left panel}: fiber-optics components. Superluminescent diode (SLED) at 1550 nm with 50 nm bandwidth, single mode fiber (light blue), isolator, circulator, polarization--maintaing (PM) fiber (dark blue), polarizer along slow axis of the fiber and at 45 degrees between slow and fast axis, electro--optical modulator (EOM), 5 meter fiber patch, photodetector with 125 MHz bandwidth and AC coupled output, lock--in amplifier. In practice, circulator with fast--axis blocking is used to eliminate the need of in--fiber polarizer, and input port of EOM is aligned at 45${}^o$ with respect to fiber axis, which is schematically represented as stand--alone 45${}^o$ polarizer, which is likewise not present in our apparatus.
        \textbf{Right panel}: schematics of free--space components and cryogenic parts. Quarter--wave plate (QWP), lens with 50 mm focal length, fused silica window at top of the cryostat, coil magnet, sample glued to copper cold--finger.
    }
\label{fig:ZALSI}
\end{figure*}

\section{Zero-area-loop Sagnac interferometer}\label{app:ZALSI}
\setcounter{equation}{0}\renewcommand{\theequation}{A.\arabic{equation}}

Design of fiber--optic zero loop area fiber Sagnac interferometer (ZALSI) \cite{Xia2006} used in the present study is schematically illustrated on \figref{fig:ZALSI}. Continuous wave superluminescent diode (SLED) with 1550 nm center wavelength is used to emit light into polarization--maintaining (PANDA) fiber, which is split equally between slow and fast axis of the fiber using a polarizer oriented at 45${}^o$ degrees. Due to finite bandwidth of about $50$ nm the SLED spectrum, two beams propagating along orthogonal axis of the fiber can be considered non-coherent after traveling for just $10$ cm along the PM fiber with birefringence $\Delta n = 4 \times 10^{-4}$. Each light wave--pocket passes through the quarter--wave plate and is converted to right and left circularly polarized beams (which do no add up to a linearly polarized light since they are decoherent). Upon reflection from the sample, say, right--circularly polarized $\vc{E} = E_0e^{-ikz}(\hat{x} + i\hat{y})$ light will acquire phase and amplitude change different from left--circularly polarized light \emph{only if} time-reversal symmetry is broken in the material \cite{Halperin1992}. Upon reflection polarization of the light is fully converted to left--circularly polarized light $\vc{E}' = E_0e^{ikz}(\hat{x} + i\hat{y})e^{-i\theta_K}$ which after passing quarter--wave plate goes into complimentary fast axis of the fiber. Similarly, light beam traveling along the fast--$y$ axis to the sample, propagates along the slow--$x$ axis back from the sample. Once the beams reach the 45${}^o$ degree again, they have traveled the same optical length, hence they interfere, and interference pattern is sensitive to $2\theta_K$ phase shift between the beams.

Importantly, our apparatus is robust against any time--reversal preserving optical activities such linear birefringence or chirality, because these activities would only result in a mix of right and left circularly polarized (elliptically polarized) light after reflection, but would not change the interference pattern at the detector. Indeed, if, say some part of the light beam traveling along the slow axis of the fiber gets reflected back into the slow axis, by the time it reaches the detector it would have acquired such a phase $\varphi_{xx} = 2n_\text{slow}L$, where $L\simeq5$ m is the distance between the 45${}^o$ degree polarizer and the sample. Only the beams traveling along the time-reversal symmetric copies of each others path and acquiring phase $\varphi_{xy} = n_\text{slow}L + n_\text{fast}L$ will interfere as guaranteed by finite bandwidth of the source and a long 5 m fiber patch.

In order to measure phase shift $2\theta_K$ induced by magneto--optical Kerr effect, we use an electro--optical modulator (EOM) to modulate the phase of the light passing through the slow axis of the fiber (see \figref{fig:ZALSI}). As a result, it can be shown using straightforward Jones analysis of polarization state of the light along the beam path, that for an arbitrary sample described by Jones matrix
\begin{align}
    J_\text{sample} = \begin{pmatrix}
        R_+	&	R_{\mp}	\\
        R_{\pm}	&	R_-
    \end{pmatrix}
\end{align}
power at the detector (up to an overall factor) is given by
\begin{align}\label{eq:power_at_detector}
    P &= |R|^2 + 2|R_+R_-|\cos2\left[\theta_K - \phi_m\sin(\omega t + \delta)\right],
\end{align}
where first term $|R|^2 \equiv |R_{+}|^2 + |R_{\pm}|^2 + |R_{\mp}|^2 + |R_{-}|^2$ is the total power without modulation and interference, while the second term comes from interfering two counter--propagating beams.  
Phase shift $\delta = (\omega-\omega_p)\frac{\tau}{2}$ is the phase acquired from traveling from EOM to the sample and back. We fix modulation frequency to a proper value $\omega_p = \frac{\pi}{\tau} \simeq 10$ MHz determined by the time $\tau$ it takes the light to make the round trip. In doing so we minimize contribution from the residual--amplitude modulation, which is ignored in the formulas above and below. Producing Fourier decomposition of signal \eqref{eq:power_at_detector} we split the signal into three main components:
\begin{align}
    P_0' & = \frac{|R|^2}{2|R_{+}R_{-}|} + J_0(2\phi_m),	\\
    P_{\omega}' &= 2 \sin (2\theta_K) J_1(2\phi_m) \sin(\omega t + \delta), \\
    P_{2\omega}' &= 2 \cos (2\theta_K) J_2(2\phi_m) \cos(2\omega t + 2\delta).
\end{align}
where $J_1$ and $J_2$ are Bessel functions and we re-scaled power 
$$P' = \frac{P}{2|R_+R_-|}$$
for brevity. From here we see that first harmonic $P_\omega$ is only present when $\theta_K \neq 0$, and Kerr angle can be calculated as
\begin{equation}\label{eq:Sagnac_kerr}
    \theta_K = \frac{1}{2} \tan^{-1}\left[ \frac{J_2(2\phi_m)V_{\omega}^\text{RMS}}{J_1(2\phi_m) V_{2\omega}^\text{RMS}}\right],
\end{equation}
which is independent of optical power or sample reflectivity. Here $V_{\omega}^\text{RMS} \propto P_\omega$ is the voltage reading on the lock--in amplifier.
Finally, to maximize signal--to--noise ratio we set modulation amplitude value $\phi_m$ and first lock--in phase $\delta$ such that first harmonic is maximized when measured from a magnetic material.

Our apparatus has proven itself an extremely sensitive magneto--optical measurement technique, which agrees with other Kerr angle probes when tested on a strongly magnetic materials, and at the same time is able to detect much smaller signals unreachable via different MOKE experimental tests, as it is evident from the recent study of charge order in CsV$_3$Sb$_5$ \cite{Saykin2023,Farhang2023}. Due to remarkably high shot--noise--limited sensitivity of $100$ nrad/$\sqrt{\text{Hz}}$ zero loop area Sagnac interferometer is even able to observe a time-reversal breaking superconducting order parameter in heavy--fermion materials Sr$_2$RuO$_4$ \cite{Xia2006Sr2RuO4}, UPt$_3$ \cite{Schemm2014} and UTe$_2$ \cite{Hayes2021,Wei2022}, as well as many other phenomena.

\begin{figure}[h]
    \centering
    \includegraphics[width=.6\linewidth]{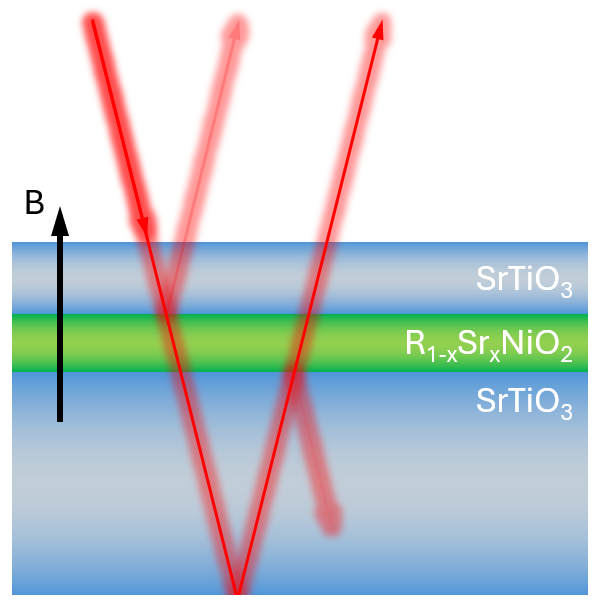}
    \caption{
        \textbf{Schematic propagation of light.} Reflected light is a combination of beams reflected different surfaces.
    }
    \label{fig:light_propagation}
\end{figure}
\begin{figure}[h]
    \centering
    \includegraphics[width=\linewidth]{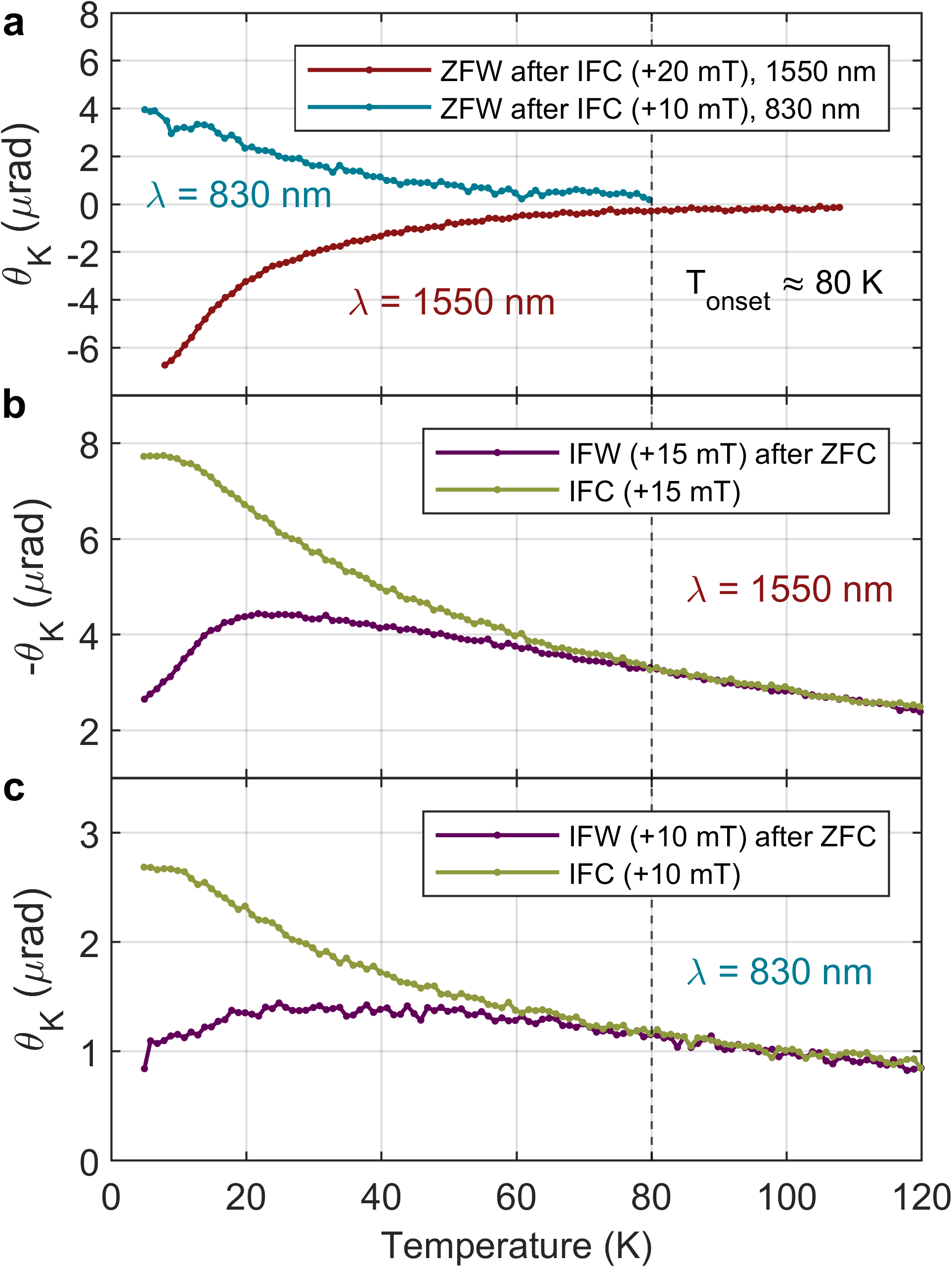}
    \caption{
        \textbf{Wavelength dependence of MOKE.} 
        Kerr signal in \LSNO vs temperature measured with two different ZALSI interferometers operating at 1550 and 830 nm.
        \textbf{a} Zero-field warmup after training in positive field at two different wavelengths.
        \textbf{b} In-field warmup/cooldown measured at 1550 nm. \textbf{c} In-field warmup/cooldown measured at 830 nm.
    }
    \label{fig:MOKE_wavelength}
\end{figure}

\section{Relationship between MOKE and magnetization}
Light beam experiences several reflections upon its optical path on each interface. However STO capping layer and substrate are transparent, while nickelate layer is mostly transparent. It has resistivity of $\rho \simeq 200$ $\mu$Ohm*cm which correspond to skin--depth $\delta \simeq 50$ nm, much larger then sample's thickness of $d \simeq 5$ nm. Overall, returned light is mostly composed of the light reflected from the bottom surface of the substrate with small contribution from the metallic surface (see Fig.\figref{fig:light_propagation}). As a result (in the absence of magnetic field) measured optical rotation consists mostly of combination of two Faraday angle acquired upon propagation in two different directions.
\begin{align}
    \theta_K^\text{total} \simeq \theta_F^\swarrow + \theta_F^\nearrow.
\end{align}
It is important to highlight that Faraday effect measured in transmission only can be present due to non--magnetic types of optical activities, such as birefringence or dichroism, however, a combination of two counter--propagating Faraday rotations has the same symmetry as a Kerr angle \cite{Halperin1992}.
Situation is more complicated, however, in the presence of external magnetic field, since then the substrate contributes to the optical rotation as well.

As follows directly from Fresnel equations, Kerr and Faraday rotations are given by \cite{Oppeneer1999}
\begin{align}
    \theta_K &= \operatorname{Re}\left\{ \frac{\sigma_{xy}}{\sigma_{xx} \left[ 1 + 4\pi i \sigma_{xx}/\omega \right]^{1/2}} \right\}, \\
    \theta_F &= \frac{2\pi d}{c}\operatorname{Re}\left\{ \frac{\sigma_{xy}}{ \left[ 1 + 4\pi i \sigma_{xx}/\omega \right]^{1/2}} \right\},
\end{align}
where $d$ is the thickness of the material. The meaning behind this equation is simple: when electromagnetic wave propagates inside the media, it constantly "shakes" electrons, which in turn re--emitted the light as radiation, however in the presence of magnetic moment, electron's orbital motion is curved (described by finite $\sigma_{xy}(\omega)$), which finally leads to rotation of polarization. Such optical activity is known as gyrotropy and is characteristic to many magnetic materials \cite{Landau2013}. What is not generally simple is a relation between Hall conductivity $\sigma_{xy}(\omega)$ and magnetization $M_z$ at optical frequencies, since it's determined by the exact structure of all the allowed optical transitions. In general, Hall conductivity $\sigma_{xy}(\omega)$ typically changes sign as a function of light energy (wavelength).

\begin{figure}[h]
    \centering
    \includegraphics[width=\linewidth]{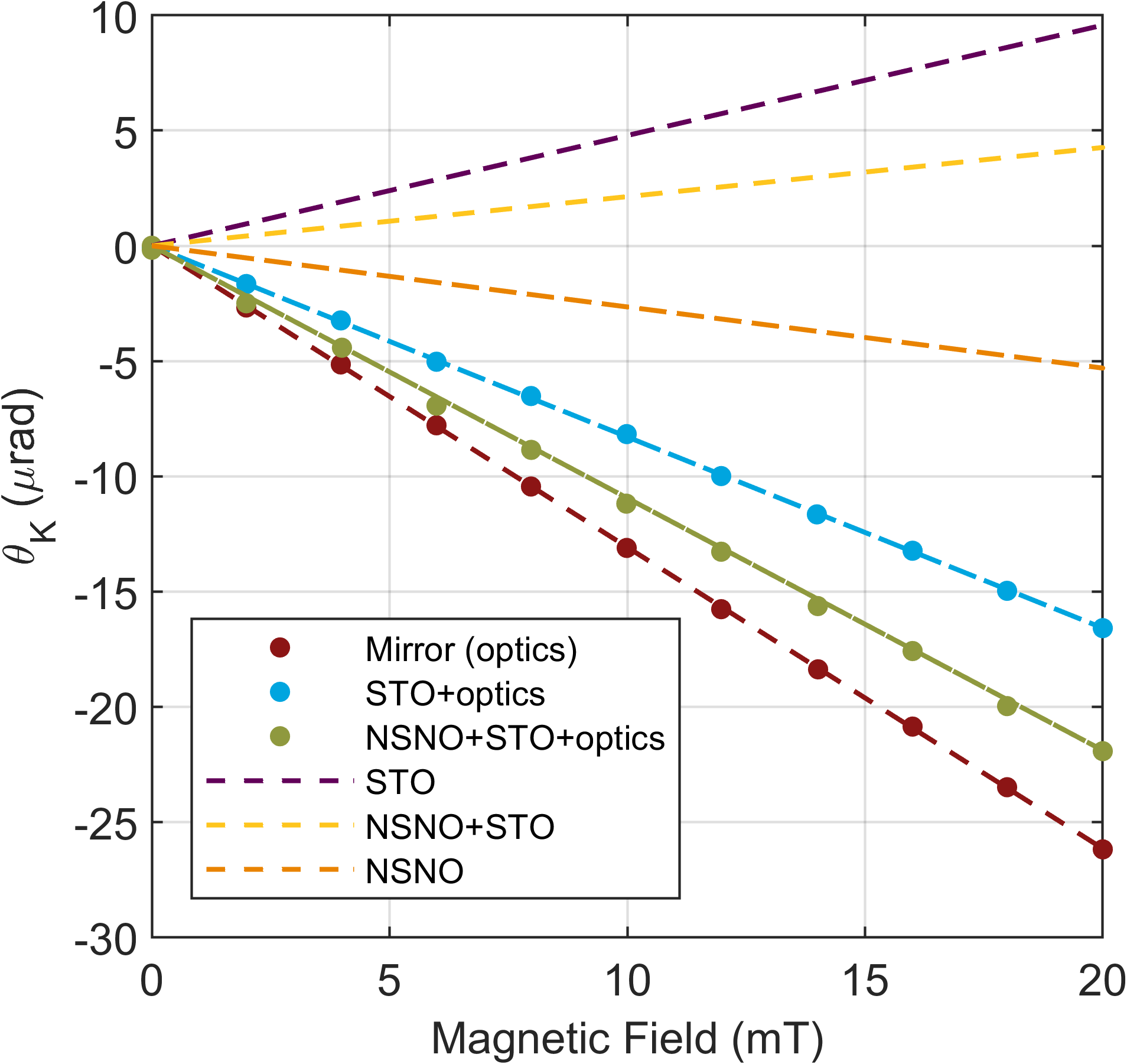}
    \caption{
        \textbf{MOKE vs Magnetic Field.} Kerr signal as a function of applied magnetic field at $T=150$ K. Circle points represent measured data and dashed lines correspond to linear fit.
    }
    \label{fig:field_sweep}
\end{figure}

\begin{figure}[h]
    \centering
    \includegraphics[width=\linewidth]{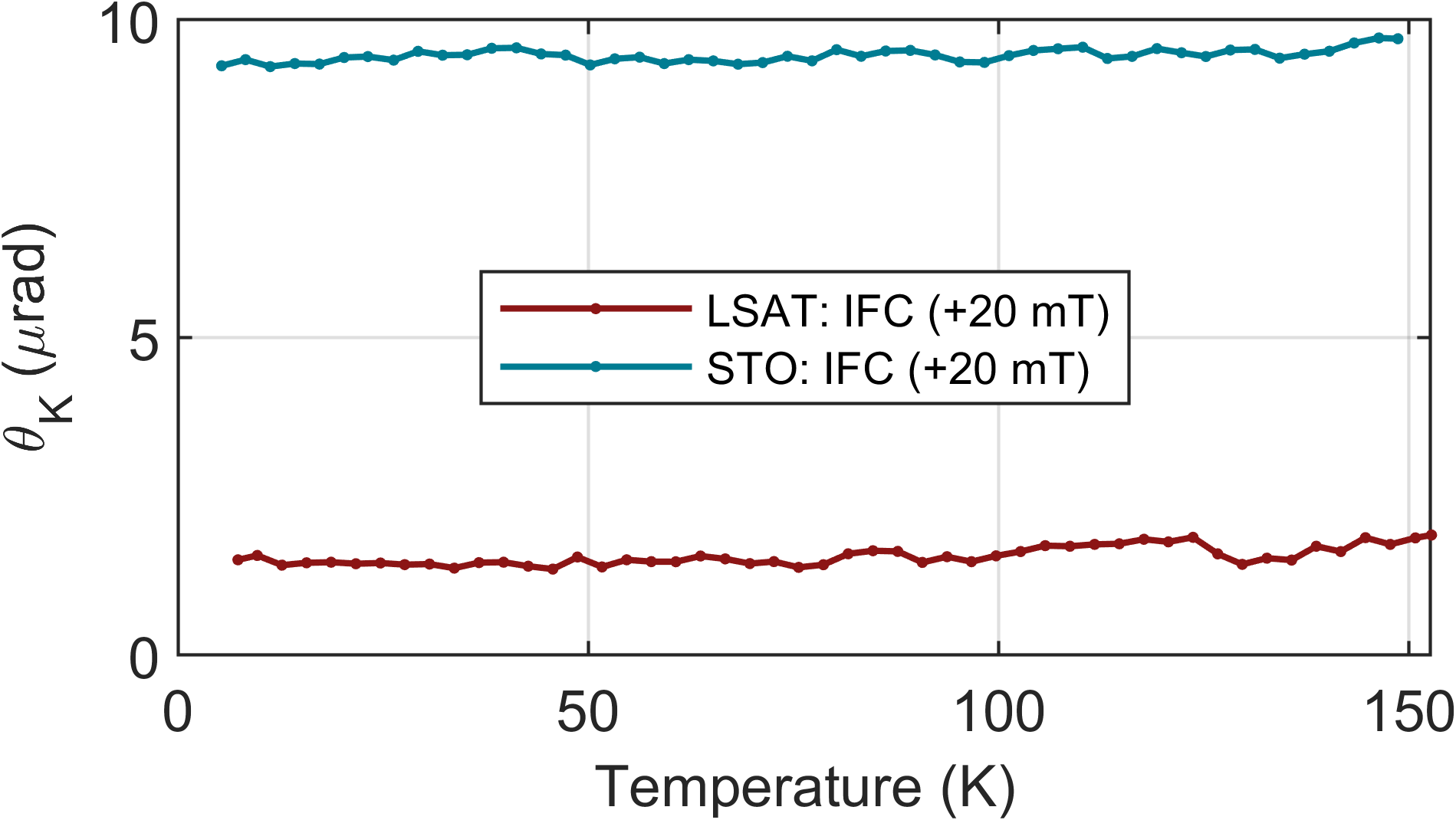}
    \caption{
        \textbf{MOKE in STO and LSAT.} Paramagnetic--like Kerr signal in \STO and \LSAT substrates measured during cooldown in $B=+20$ mT magnetic field. Contribution from optics was subtracted.
    }
    \label{fig:MOKE_STO_LSAT}
\end{figure}

\section{In--field MOKE measurements}\label{app:in_field}
All measurements performed in the presence of magnetic field suffer from background signal caused by Faraday effect in optical elements along light's path: quarter--wave plate, lense, and cryostat window all contribute to measured MOKE value. This additional wavelength--dependent linear--in--field part of the signal can be estimated based on Verdet constant of the material, thickness of optical elements and magnetic field they experience. However, in practice magnetic field varies with distance from the cryostat, which results in noticable change in measured signal even when lens is moved by 1 mm, so it is easier to measure contribution from optics by placing a mirror (non--magnetic reflecting material) next to the sample and measuring signal from it.

Results of such calibration measurement performed on NSNO--on--STS sample, STO substrate, and mirror are shown on \figref{fig:field_sweep}. Magnetic field sweeps were performed at $T=$ 150 K, where no history--depended spin--glass dynamics is expected to be present. Kerr rotation measured off of the mirror, comes fully from the Faraday rotation happening in the optical elements, and is as expected negative in positive field, due to negative Verdet constant of fused silica glass $V_\text{glass} \simeq 2.5$ rad$\cdot$T${}^{-1}$$\cdot$m${}^{-1}$ at 1550 nm. 
Measurement of \STO substrate combine a combination of positive paramagnetic--like contribution from the substrate and negative signal from the optics, subtracting value recorded off of mirror, we can isolate Kerr rotation produced by \STO substrate. 
Finally, to separate signal from the nickelate we look at the difference between Kerr signal detected from the sample and from the substrate.

Furthermore, we repeat substrate measurement at fixed applied magnetic field of $+20$ mT and sweep the temperature to obtain data presented on \figref{fig:MOKE_STO_LSAT}. We observe almost no temperature dependence of STO--induced Kerr signal, which lets us conclude that observed glassy phase (Fig.~1 of the main text) is coming fully from the sample.

Lastly, let us emphasize that unlike in--field measurement, running experiment in the at $B = 0$ mT has no such drawbacks, hence there is no need to perform a background subtraction for the zero--field warmup datasets.


\end{document}